%% file: main.tex
\documentclass[acmsmall]{acmart}

\makeatletter
\@ifpackageloaded{amssymb}{}{}
\makeatother

\usepackage{graphicx}
\usepackage{booktabs}
\usepackage{multirow}
\usepackage{tcolorbox}
\usepackage{xcolor}
\usepackage{listings}
\usepackage{tabularx}

\definecolor{darkblue}{rgb}{0.0, 0.2, 0.6}
\definecolor{entitycolor}{RGB}{139,0,0}
\definecolor{entitytypecolor}{rgb}{0.0, 0.2, 0.6}
\definecolor{intentnamecolor}{RGB}{139,0,0}
\definecolor{keywordcolor}{rgb}{0.0, 0.2, 0.6}
\definecolor{textcolor}{rgb}{0.1, 0.1, 0.1}

\tcbset{arc=3pt, boxrule=\heavyrulewidth, colback=gray!5, colframe=black, boxsep=0pt}

\newcommand{\header}[1]{\smallskip\noindent\textbf{#1}}

\newcommand{\RQI}{How well do the generated LFs label data?}
\newcommand{\RQII}{What characteristics impact the performance of the generated LFs?}

\begin{document}

\title{An Approach for Auto Generation of Labeling Functions for Software Engineering Chatbots}

\author{Ebube Alor}
\affiliation{%
  \department{Data-driven Analysis of Software (DAS) Lab}
  \department{Department of Computer Science \& Software Engineering}
  \institution{Concordia University}
  \city{Montréal}
  \state{QC}
  \country{Canada}}
\email{ebubechukwu.alor@mail.concordia.ca}

\author{Ahmad Abdellatif}
\affiliation{%
  \department{Department of Electrical \& Software Engineering}
 \institution{University of Calgary}
 \city{Calgary}
 \state{AB}
 \country{Canada}}
\email{ahmad.abdellatif@ucalgary.ca}

\author{SayedHassan Khatoonabadi}
\affiliation{%
  \department{Data-driven Analysis of Software (DAS) Lab}
  \department{Department of Computer Science \& Software Engineering}
  \institution{Concordia University}
  \city{Montréal}
  \state{QC}
  \country{Canada}}
\email{sayedhassan.khatoonabadi@mail.concordia.ca}

\author{Emad Shihab}
\affiliation{%
  \department{Data-driven Analysis of Software (DAS) Lab}
  \department{Department of Computer Science \& Software Engineering}
 \institution{Concordia University}
 \city{Montréal}
 \state{QC}
 \country{Canada}}
\email{emad.shihab@concordia.ca}

\renewcommand{\shortauthors}{Alor et al.}

\begin{abstract}
    Software engineering (SE) chatbots are increasingly gaining attention for their role in enhancing development processes. At the core of chatbots are Natural Language Understanding platforms (NLUs), which enable them to comprehend user queries but require labeled data for training. However, acquiring such labeled data for SE chatbots is challenging due to the scarcity of high-quality datasets, as training requires specialized vocabulary and phrases not found in typical language datasets. Consequently, developers often resort to manually annotating user queries---a time-consuming and resource-intensive process. Previous approaches require human intervention to generate rules, called labeling functions (LFs), that categorize queries based on specific patterns. To address this issue, we propose an approach to automatically generate LFs by extracting patterns from labeled user queries. We evaluate our approach on four SE datasets and measure performance improvement from training NLUs on queries labeled by the generated LFs. The generated LFs effectively label data with AUC scores up to 85.3\% and NLU performance improvements up to 27.2\%. Furthermore, our results show that the number of LFs affects labeling performance. We believe that our approach can save time and resources in labeling users' queries, allowing practitioners to focus on core chatbot functionalities rather than manually labeling queries.
\end{abstract}




\maketitle

\section{Introduction}
\label{sec:introduction}
\input{Introduction}

\section{Background}
\label{sec:background}
\input{Background}

\section{Approach}
\label{sec:approach}
\input{Approach}

\section{Case Study Setup}
\label{sec:study_setup}
\input{Studysetup}

\section{Results}
\label{sec:results}
\input{Results}

\section{Discussion}
\label{sec:discussion}
\input{Discussion}

\section{Threats to Validity}
\label{sec:threats}
\input{Threats}

\section{Related Work}
\label{sec:related_work}
\input{Relatedwork}

\section{Conclusion}
\label{sec:conclusion}
\input{Conclusion}

\bibliographystyle{ACM-Reference-Format}
\bibliography{references}

\clearpage
\appendix
\label{sec:appendix}
\input{Appendix}

\end{document}

%% file: Introduction.tex
In the field of software engineering (SE), chatbots are deployed as conversational tools to automate a wide range of tasks, from assisting in problem-solving to answering questions about repositories~\cite{grudin2019chatbots, ross2023programmer, tony2022conversational}. At the core of these chatbots are Natural Language Understanding platforms (NLUs). These NLUs serve as the backbone of the chatbot, responsible for interpreting human language into structured data that chatbots can act on~\cite{abdellatif2021comparison}. They accomplish this by using machine learning and natural language processing techniques to dissect user queries, identifying the user's intent and extracting key entities like bug IDs or version numbers~\cite{prajwal2019universal}. By doing so, NLUs make it possible for the chatbot to generate responses that are contextually relevant and specific to the user's queries.

The effectiveness of NLUs heavily relies on the availability of a large volume of high-quality training data~\cite{zarcone2021small, lin2023review, parrish2021does}. However, prior work shows that obtaining such data is expensive, especially in specialized domains like SE~\cite{lin2020msabot,Dominic2020ICSEW,abdellatif2020msrbot}, where the domain-specific language and terminology pose unique challenges for data collection and labeling~\cite{abdellatif2021comparison}. For example, common words like `push', `fork', and `commit' have distinct meanings in the context of software development, which differ from their everyday usage~\cite{abdellatif2021comparison}. This makes it difficult to rely on general-purpose datasets for training SE chatbots. Furthermore, there is a notable scarcity of publicly available, high-quality datasets for training chatbots in the SE domain~\cite{motger2021conversational}. As a result, gathering sufficient and relevant training data for SE chatbots often requires significant effort and resources.

To obtain the necessary training data for SE chatbots, practitioners often resort to mining chatbot user queries as a data source~\cite{farhour2024weak}. However, chatbot developers need to label these queries before they can be used for training the NLU, introducing significant challenges~\cite{li2017dailydialog}. First, manual labeling is labor-intensive and can incur substantial costs~\cite{tu2020better}, particularly when domain expertise is essential for ensuring the accuracy of labels~\cite{pinhanez2019machine, zhang2016learning}. This increases the financial burden and extends the time required to prepare data for training purposes~\cite{ghahreman2011semi, tu2020better, pujara2011reducing}. Second, although semi-automated labeling methods exist, they still require human intervention, such as domain experts developing heuristics for labeling user queries~\cite{galhotra2021adaptive, li2021weakly, hancock2018training}. These heuristics are also called Labeling Functions (\textit{referred hereafter simply as LFs}). Specifically, LFs are programmatic rules or functions that assign labels to data points based on certain conditions or patterns. For example, an LF for an SE chatbot might label a query as `bug-related' if it contains keywords like `error', `issue', or `fix'.

In response to the challenges outlined above regarding data labeling, we introduce our approach that automates the process of generating LFs specifically for SE datasets. To accomplish this, our approach takes a small set of labeled data as input and automatically analyzes and extracts patterns from it, and uses those patterns to generate the LFs capable of auto-labeling data. After the generation of LFs, we use them to label data and train the NLU of the chatbot on the trained data~\cite{bocklisch2017rasa}. Our approach is organized into three main components. First, the \textbf{Grouper} is responsible for expanding the initial labeled data by identifying similar queries~\cite{schopf2023semantic}. Second, the \textbf{LF Generator} takes on the role of extracting patterns from this expanded data to create LFs~\cite{bringer2019osprey}. Finally, the \textbf{Pruner} filters out low-quality LFs from the pool of generated LFs~\cite{varma2018snuba, ratner2020snorkel}. We evaluate our approach based on four datasets (namely, AskGit, MSA, Ask Ubuntu, and Stack Overflow) used to develop chatbots for performing various SE tasks. Specifically, we aim to answer the following research questions in this paper:

\smallskip
\begin{itemize}
    \item[\textbf{RQ1:}] \textbf{\RQI} We evaluate the quality of the generated LFs in terms of their effectiveness in labeling data, and subsequently, in training SE chatbots~\cite{yang2022learning, ferri2009experimental}. Our analysis shows that the generated LFs effectively label data with AUC scores of above 75.5\% for three out of the four studied datasets (except Stack Overflow). Additionally, we observe that for these three datasets, using the auto-labeled data for training can enhance the NLU's performance, with AUC score improvements of up to 27.2\%.
    \item[\textbf{RQ2:}] \textbf{\RQII} We investigate the specific characteristics of the generated LFs (i.e., coverage, accuracy, and LF support) that contribute to the LFs performance. Our findings indicate that higher values in these LF characteristics generally correlate with improved labeling performance. For instance, high coverage LFs achieve AUC scores of up to 88.3\%, compared to 50.5\% for low coverage LFs. While the influence of characteristics on performance varies, focusing solely on one characteristic may negatively impact others, suggesting a balanced approach that considers all characteristics is essential for optimal performance.
\end{itemize}

Finally, we discuss the impact of varying the number of LFs on the labeling performance. We find that a higher number of generated LFs tends to improve labeling performance. However, the rate of improvement varies across different datasets, highlighting the influence of dataset-specific characteristics on the effectiveness of the generated LFs. 

\header{Our Contributions.}
In summary, our paper makes the following contributions:
\begin{itemize}
    \item We introduce an end-to-end approach to automatically generate LFs, facilitating the training of SE chatbots.
    \item We show the effectiveness of our approach on multiple SE datasets and with the Rasa NLU platform.
    % \item We discuss the characteristics and the impact of various configurations on the quality of generated LFs.
    \item We discuss the impact of varying the number of LFs on the labeling performance.
    \item We make our dataset and code publicly available to facilitate future research in this area~\cite{chatmentrepo}.
\end{itemize}

\header{Paper Organization.}
The remainder of the paper is organized as follows. Section~\ref{sec:background} provides the background that forms the basis for our study. Section~\ref{sec:approach} presents the details of our approach and its key components. Section~\ref{sec:study_setup} describes the setup for our empirical study followed by Section~\ref{sec:results} which presents the results of our RQs. Section~\ref{sec:discussion} discusses the relevance of NLU-based chatbots in the era of LLMs and analyzes the impact of the number of LFs on labeling performance. Section~\ref{sec:threats} discusses threats to the validity of our study. Finally, Section~\ref{sec:related_work} reviews the related work and Section~\ref{sec:conclusion} concludes the paper.

%% file: Background.tex
Before proceeding to our approach, we explain in this section the chatbot-related terminologies, data labeling process, and LFs. Also, we discuss the role of NLU-based chatbots in the era of LLMs.

\subsection{Chatbot Training and Data Challenges}
\label{sec:motivation}
Software chatbots serve as the conduit between users and services~\cite{lebeuf2017software}. Users input their queries to the chatbot in natural language, which then performs the requested action (e.g., querying the database) and responds to the user's question. NLU platforms are the backbone for chatbots to understand the user's question. Specifically, NLU extracts two key aspects from the input query: the topic of the user's query or what is being asked about (known as the `Entity') and what the user is asking it to do (known as the `Intent').

Similar to any machine learning model, NLUs need to be trained to extract intents and entities. In particular, for each intent, the NLU needs to be trained on a set of queries that represent different ways a user could express that intent. Therefore, the training data should include a wide range of examples for each intent to capture the variety of ways in which people express their needs. Figure~\ref{fig:training_data_sample} presents a snapshot of training data for the \textit{FileCreator} and \textit{IssueClosingDate} intents with their training examples. Chatbot practitioners need to brainstorm the different ways that the user could ask the question for a specific intent~\cite{abdellatif2020msrbot} to train the NLU. Nevertheless, this is a resource-intensive and time-consuming task~\cite{farhour2024weak}.

\begin{figure}[b]
\centering
\begin{minipage}{0.93\columnwidth}
\lstdefinestyle{yaml}{
    basicstyle=\footnotesize\ttfamily,
    % numbers=left,
    frame=trBL,
    frameround=fttt,
    showstringspaces=false,
    breaklines=true,
    morecomment=[l]{//},
    morestring=[b]",
    morestring=[d]',
    escapeinside={(*@}{@*)},
    tabsize=1
}

\begin{lstlisting}[style=yaml]
 1 - (*@\textcolor{keywordcolor}{intent}@*): (*@\textcolor{intentnamecolor}{FileCreator}@*)
 2   (*@\textcolor{keywordcolor}{examples}@*): | 
 3    - (*@\textcolor{textcolor}{I want to know who created file}@*) 
 4      [(*@\textcolor{entitycolor}{map.json}@*)] (*@\textcolor{entitytypecolor}{(file\_name)}@*)
 5    - (*@\textcolor{textcolor}{I would like to know who first created}@*) 
 6      (*@\textcolor{textcolor}{file}@*) [(*@\textcolor{entitycolor}{tf\_env\_collect.sh}@*)](*@\textcolor{entitytypecolor}{(file\_name)}@*)?
 7    - (*@\textcolor{textcolor}{creator of file}@*) [(*@\textcolor{entitycolor}{server.json}@*)](*@\textcolor{entitytypecolor}{(file\_name)}@*)
 8    - (*@\textcolor{textcolor}{developer who created}@*) [(*@\textcolor{entitycolor}{constant.txt}@*)]
 9      (*@\textcolor{entitytypecolor}{(file\_name)}@*)
10 - (*@\textcolor{keywordcolor}{intent}@*): (*@\textcolor{intentnamecolor}{IssueClosingDate}@*)
11   (*@\textcolor{keywordcolor}{examples}@*): |
12    - (*@\textcolor{textcolor}{when was}@*) [(*@\textcolor{entitycolor}{issue 24}@*)](*@\textcolor{entitytypecolor}{(issue\_number)}@*) (*@\textcolor{textcolor}{closed}@*)
13    - (*@\textcolor{textcolor}{show me the closing date for}@*) [(*@\textcolor{entitycolor}{issue 3}@*)]
14      (*@\textcolor{entitytypecolor}{(issue\_number)}@*)
15    - (*@\textcolor{textcolor}{What was the date of closing}@*) [(*@\textcolor{entitycolor}{issue 4}@*)]
16      (*@\textcolor{entitytypecolor}{(issue\_number)}@*)
17    - (*@\textcolor{textcolor}{When was}@*) [(*@\textcolor{entitycolor}{issue 1109}@*)](*@\textcolor{entitytypecolor}{(issue\_number)}@*) 
18      (*@\textcolor{textcolor}{closed}@*)
    
\end{lstlisting}
\end{minipage}
\caption{Example dataset for training an SE chatbot.}
\label{fig:training_data_sample}
\end{figure}

\begin{figure}
	\centering
	\hspace{1\linewidth}
	\includegraphics[width=0.55\linewidth]{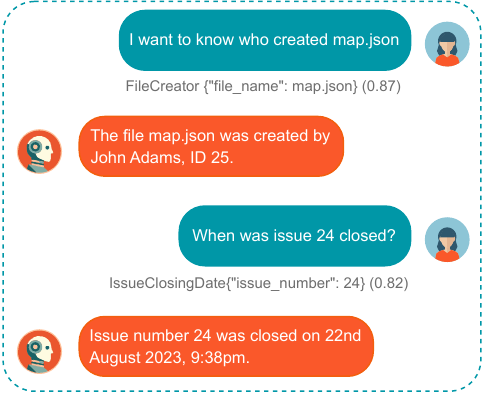}	
	\caption{Example of a user's interaction with an SE chatbot.}
	\label{fig:chatbot_interaction}	
\end{figure}

Alternatively, chatbot practitioners leverage user-chatbot conversations to augment their dataset~\cite{farhour2024weak}. To demonstrate this, we present the user-chatbot interaction example shown in Figure~\ref{fig:chatbot_interaction}. In the example, the user (highlighted in green) asks the chatbot software repository-related questions. For the first query, the user asks, `I want to know who created map.json'. Here, the intent is to find out the developer who created the file. The chatbot's NLU analyzes this query, extracting the intent (\textit{FileCreator} with a confidence score of 0.87) and the relevant entity (\textit{file\_name} is map.json). Based on this understanding, the chatbot then provides a response identifying the file creator.

To continuously improve the chatbot's performance, developers use these user-chatbot conversations to expand the NLU's training data. They review queries, especially those with low confidence scores, validate the extracted intent, make corrections if necessary, and then add these annotated queries to the training dataset. Although this process might require less time compared to brainstorming new queries for intent, the chatbot developer still requires annotating users' queries to improve the NLU's performance. To reduce the burden and save developers time, we propose an approach that automates the annotation process of the users' queries.

\subsection{Data Labeling}
The process of data labeling involves assigning appropriate labels to a dataset~\cite{ratner2020snorkel, galhotra2021adaptive}. There are several data labeling techniques that are traditionally used~\cite{prenner2021making, robbes2019leveraging}. These include manual labeling, crowd-sourcing, and hiring domain experts~\cite{prenner2021making, tu2020better, pinhanez2019machine, foncubierta2012ground}.

Manual labeling is the annotation of unlabeled data by humans~\cite{li2017dailydialog}. The process typically begins with gathering the unlabeled data~\cite{hancock2019learning} and reviewing them. Next, appropriate labels for intents and entities are assigned to the unlabeled data, categorizing them into classes~\cite{hancock2019learning}. Once the data is suitably annotated, it is incorporated into the chatbot's training dataset to further enhance its accuracy~\cite{hancock2019learning}. Because of all these steps, the manual labeling process requires significant effort and is time-consuming~\cite{tu2020better}. It can also lead to inconsistencies due to human error, making it less practical for large-scale projects~\cite{parrish2021does}. 

Crowdsourcing involves distributing the task of labeling data to a large group of people, often through an online platform. This method can significantly accelerate the data labeling process by leveraging the crowd's collective effort. Although crowdsourcing can help label a large volume of data quickly, it may not always guarantee the quality necessary for specialized domains like SE. The lack of domain-specific knowledge among the crowd can lead to inaccuracies in labeling, affecting the overall quality of the training data~\cite{zhang2016learning, sheng2019machine, stolee2010exploring}.

Recruiting domain experts refers to engaging individuals with specialized knowledge in a particular area, such as SE, to label the data. These experts bring a high level of accuracy and insight to the labeling process, ensuring the data is correctly annotated with the appropriate intents and entities. However, this method can be costly and may not scale well for large datasets, making it a challenging option for projects with limited budgets~\cite{pinhanez2019machine, zhang2016learning}. Due to the limitations of these labeling techniques, an alternative approach is for domain experts to craft rules or heuristics that can then be applied to unlabeled data~\cite{ratner2017snorkel}. These rules are called Labeling Functions (LFs) and will be discussed in the next subsection.

\subsection{Weak Supervision and LFs}

Weak supervision is a machine learning approach that addresses the challenge of obtaining large amounts of accurately labeled data~\cite{ratner2017snorkel, karamanolakis2021self}. Instead of relying solely on expensive and time-consuming manual annotation, weak supervision leverages noisy, imprecise, or incomplete sources of information to generate training labels~\cite{karamanolakis2021self, boecking2020interactive}. Different forms of weak supervision exist~\cite{hernandez2016weak}, including:
\begin{itemize}
    \item \textbf{Heuristics:} Rule-based labeling using domain knowledge~\cite{boecking2020interactive} (e.g., labeling a query as `bug-related' if it contains the word `error').
    \item \textbf{Distant Supervision:} Using an external knowledge base or database to automatically generate labels~\cite{mintz2009distant} (e.g., labeling code comments based on the presence of specific API calls).
    \item \textbf{Third-party Models:} Using pre-trained models to generate labels for new data, even if the models were trained on different but related tasks~\cite{mann2010generalized}.
\end{itemize}

The key advantage of weak supervision is its ability to significantly reduce the time and cost associated with data labeling while enabling the use of larger datasets~\cite{ratner2017snorkel}. This can lead to the development of more robust and accurate models. Moreover, weak supervision, especially through the use of heuristics, allows for the direct integration of valuable domain expertise into the labeling process~\cite{varma2018snuba}.

LFs are a common and practical way to implement weak supervision. They are essentially a set of heuristic rules or predefined conditions to assign labels to data~\cite{ratner2017snorkel, lison2021skweak, ratner2020snorkel}. Instead of manually labeling each piece of data, a domain expert formulates a set of rules or conditions to create an LF. The primary goal of LFs is to facilitate the labeling process by determining the appropriate labels for each piece of data~\cite{ratner2017snorkel, lison2021skweak, ratner2020snorkel}. LFs offer several key advantages over manual labeling, including:
\begin{itemize}
    \item \textbf{Reusability:} Once crafted, LFs can be applied multiple times across various batches of data~\cite{ratner2017snorkel, lison2021skweak}.
    \item \textbf{Efficiency:} LFs can significantly streamline the labeling process by automating label assignment based on well-defined criteria~\cite{ratner2017snorkel, lison2021skweak, ratner2020snorkel}.
\end{itemize}

Despite these advantages, developing effective LFs can be challenging, as it requires deep domain expertise and substantial time investment, especially as the complexity of the data and the rules increases~\cite{galhotra2021adaptive}. To streamline the management and application of our LFs, we selected Snorkel as the underlying framework~\cite{ratner2020snorkel}. Snorkel is a framework specifically designed to enable and manage weak supervision through the use of LFs~\cite{ratner2017snorkel}. It is developed for programmatically building and managing training datasets without manual labeling~\cite{ratner2017snorkel}. Snorkel provides a structured approach for writing LFs in Python and uses a generative model to combine their outputs to create probabilistic labels for unlabeled data. Each LF in Snorkel is a standalone function that labels a subset of the data based on a certain rule or heuristic~\cite{ratner2017snorkel}.

Consider an example of a Snorkel LF implemented in Python, the \textit{ContainsWordLabeler} shown in Figure~\ref{fig:lf_python_implementation}. It is designed to identify intents based on words a query contains. The \textit{ContainsWordLabeler} class is initialized with a dictionary of labels corresponding to different intents. The \texttt{apply} method takes the input unlabeled data and checks for the presence of phrases indicative of the \textit{FileCreator} or \textit{IssueClosingDate} intents. If such phrases are detected, it returns the associated label; if not, it returns \texttt{ABSTAIN}, indicating that the LF cannot confidently assign a label.

\begin{figure}
\centering
\begin{minipage}{0.93\columnwidth}
\lstset{
    language=Python,
    frame=trBL,
    frameround=fttt,
    basicstyle=\footnotesize\ttfamily,
    numberstyle=\tiny,
    tabsize=2,
    commentstyle=\color{gray},
    keywordstyle=\color{darkblue}
}

\begin{lstlisting}
 1 class ContainsWordLabeler(LabelingFunction):
 2     def apply(self, doc) -> int:
 3         # Phrases indicating file creator
 4         file_phrases = [
 5             'created file', 'file by', 
 6             'developer created'
 7         ] 
 8         # Phrases for issue closing date
 9         issue_phrases = [
10             'when was issue', 'issue closed'
11         ]
12         # Check for file creation phrases
13         if any(phrase in doc.text.lower() 
14                for phrase in file_phrases):
15             return self.labels['FileCreator']
16         # Check for issue closing phrases
17         if any(phrase in doc.text.lower() 
18                for phrase in issue_phrases):
19             return self.labels[
20                 'IssueClosingDate']
21         return ABSTAIN
22 
23 # Example usage with a software dataset
24 lf = ContainsWordLabeler()
25 label = lf.apply('Who made map.json?')
26 # Output is FileCreator label
\end{lstlisting}
\caption{Implementation of a Snorkel LF for identifying SE-related intents.}
\label{fig:lf_python_implementation}
\end{minipage}
\end{figure}

\subsubsection{Benefits of an LF Generation Approach}
Given the capabilities of LLMs, one might ask why our approach for \textit{generating} the LFs themselves does not primarily rely on LLMs (e.g., by prompting an LLM to write labeling rules). Our methodology, using semantic similarity (Grouper), pattern/statistic extraction and ML classifiers (Generator), and quality filtering (Pruner), was a deliberate choice based on several factors prioritizing \textbf{transparency, control, efficiency, and alignment with established weak supervision principles}:
\begin{itemize}
\item \textbf{SE Data Scarcity for LLM Fine-Tuning:} LLMs typically require substantial amounts of domain-specific labeled data for effective fine-tuning \cite{shin2023prompt}, yet such annotated datasets are particularly scarce in specialized SE contexts. Our approach addresses this fundamental challenge by automatically generating LFs that can create labeled data from readily available unlabeled queries, effectively bootstrapping the data annotation process without relying on pre-existing large-scale SE datasets.
\item \textbf{Reusability and Computational Cost:} A key output of our approach is a set of explicit LFs. These LFs are reusable artifacts; once generated, they can be applied consistently to \textbf{label future batches of unlabeled data} as they arrive \cite{ratner2017snorkel, lison2021skweak}. This contrasts with using an LLM for direct, one-time classification of a dataset, where the underlying labeling logic might not be captured as an easily reusable function for ongoing labeling needs. Reusability helps to save cost and energy resources. Also our approach utilizes relatively lightweight tools like Sentence Transformers, CountVectorizer, and standard ML libraries (e.g., scikit-learn), which are computationally less demanding and more accessible than requiring large LLMs for the core pattern-finding and rule-generation steps themselves \cite{patterson2021carbon}.
\item \textbf{Explainability and Trustworthiness:} Our generated LFs are based on identifiable features (unique words, entity types, co-occurrences) or explainable ML models. This allows practitioners to inspect \textit{why} a query receives a certain label, fostering trust and enabling easier debugging. In contrast, LLM-generated rules often function as ``black boxes'' that make verification difficult and may produce inconsistent labels for similar queries.

\end{itemize}
In essence, we traded the potential generative power of LLMs in the LF creation step for greater transparency, control, and efficiency in the labeling process itself, aligning with the practical needs of building reliable data labeling pipelines for specialized domains like SE.

%% file: Approach.tex
\begin{figure*}
	\centering
	\includegraphics[width=\linewidth]{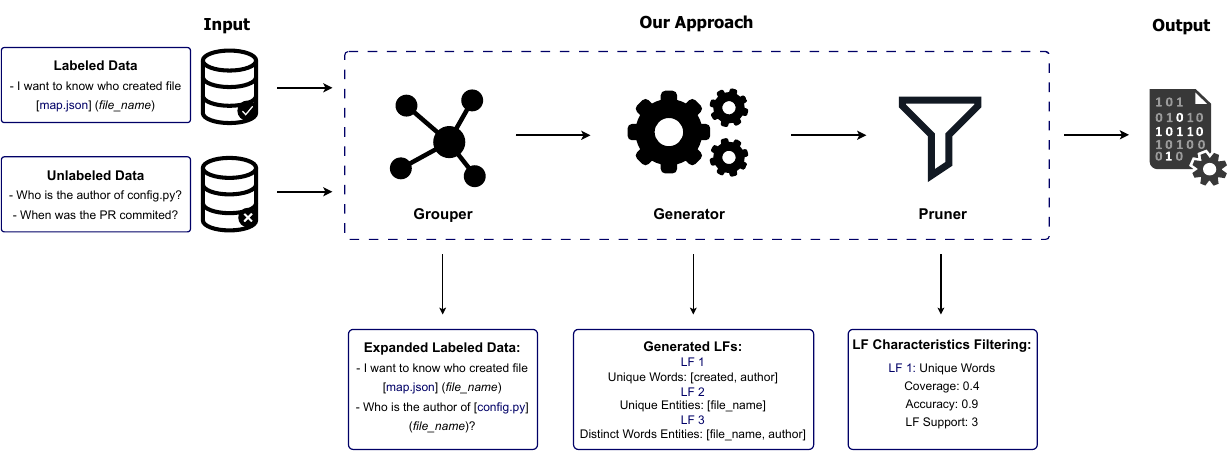}	
	\caption{Overview of our approach and its components.}
	\label{fig:approach}	
\end{figure*}

Figure~\ref{fig:approach} presents an overview of our approach, which automates the process of generating LFs. The approach takes queries along with their corresponding intents (labeled data) and queries that need to be labeled (unlabeled data) as inputs. The output of our approach is a set of generated LFs that can be used for labeling user queries. Our approach is composed of three main components: (1) Grouper, which expands the initial labeled data by identifying and incorporating semantically similar queries from the unlabeled data pool, (2) Generator, responsible for identifying and extracting patterns from the expanded labeled data to generate LFs, and (3) Pruner, which serves as a quality control filter, evaluating all candidate LFs based on their performance and discarding those that are redundant or ineffective. We detail each component in this section.

\subsection{Grouper}
\label{sec:grouper}
Prior work shows that the size and diversity of the initial labeled data directly impact the quality of the generated LFs~\cite{varma2018snuba, xu2021dp, denham2022witan}. A larger and more varied set of examples allows for the extraction of more robust and generalizable patterns. Therefore, the first component of our approach, the Grouper, is tasked with expanding the initial labeled dataset. Following established data augmentation practices~\cite{schopf2020semantic, muennighoff2023mteb}, the Grouper enriches the labeled data by incorporating semantically similar queries from the large unlabeled data pool.

To achieve this, the Grouper calculates the semantic similarity between each query in the unlabeled set and every query in the labeled seed set. If the similarity score between an unlabeled query and a labeled one exceeds a predefined threshold, the unlabeled query is assigned the intent of its most similar counterpart and integrated into the labeled data. Queries from the unlabeled set that do not meet this threshold for any intent are not grouped and are set aside for later stages. For assessing semantic similarity, we use the \texttt{Sentence-t5-xxl} model~\cite{ni2021sentencet5}, an 11-billion parameter sentence-transformer chosen for its proven effectiveness in text similarity tasks~\cite{raffel2020exploring, muennighoff2023mteb}.

The output of this component is a larger and more diverse labeled dataset, which we refer to as the \textit{expanded dataset}. This expanded dataset serves as the input for the Generator component. For instance, as illustrated in the example in Figure~\ref{fig:approach}, the Grouper identifies that the unlabeled query ``Who is the author of config.py?'' is semantically similar to the labeled query ``I want to know who created file [map.json]'', and consequently adds it to the corresponding intent class.

\subsection{Generator}
\label{sec:generator}
The Generator is the core component responsible for creating LFs. It takes the \textit{expanded dataset} from the Grouper and analyzes it to find patterns that distinguish one intent from another. Prior work shows that intents containing unique features, such as distinct words or entity types, are more easily classified by NLU platforms~\cite{abdellatif2021comparison}. Based on this, the Generator employs four distinct strategies to create a comprehensive set of candidate LFs.

To prepare the text for analysis, we first process the queries in the expanded dataset using a \texttt{CountVectorizer}, which builds a vocabulary of unigram features. To prevent data leakage, the vectorizer's vocabulary is built exclusively from this expanded dataset and is reused for all subsequent feature transformations. The entity types for all queries are extracted using the Rasa NLU platform, which has been shown to perform well on SE tasks~\cite{abdellatif2021comparison}.

\subsubsection{Exclusive Word LFs}
This strategy identifies words that are highly correlated with a specific intent. For each word in the vocabulary, we calculate an exclusivity score for each intent using Equation~\ref{eq:exclusivity}.
\begin{equation}
    Exclusivity(word, intent) = \frac{Occurrences(word, intent)}{TotalOccurrences(word)}
    \label{eq:exclusivity}
\end{equation}
A score close to 1 indicates that a word appears almost exclusively within queries for a single intent. If a word's exclusivity score for an intent exceeds a predefined threshold, an LF is generated. This LF will vote for that intent if the word is present in a query, and abstain otherwise. For instance, in the example shown in Figure~\ref{fig:approach}, the words `created` and `author` are identified as exclusive to the \textit{FileCreator} intent, resulting in the creation of LF~1.

\subsubsection{Distinct Entity LFs}
This strategy operates similarly to the word-based one but focuses on entity types. It identifies entities that are unique to a specific intent class within the expanded dataset. For each such distinct entity, an LF is created that votes for the corresponding intent whenever that entity type is detected in a query. For example, as shown in Figure~\ref{fig:approach}, an LF (LF~2) is generated to vote for an intent based on the presence of the `file\_name` entity.

\subsubsection{Distinct Word-Entity Combination LFs}
This strategy captures more complex patterns where the co-occurrence of a specific word and an entity is a strong signal for an intent, even if the individual components are not exclusive on their own. The Generator searches for these unique combinations across the dataset. If a particular word-entity pair is found to be highly indicative of a single intent, an LF is created that votes for that intent only when both the word and the entity are present. As an example, Figure~\ref{fig:approach} shows the generation of LF~3 based on the combination of the word `author` and the `file\_name` entity.

\subsubsection{Machine Learning (ML) LFs}
While heuristic-based LFs are effective, some intents are best distinguished by more complex, non-linear patterns. To capture these, we also train a suite of standard ML classifiers (Random Forest, Decision Tree, K-Nearest Neighbors, Logistic Regression, and SVM) on the vectorized features of the entire expanded dataset. Each trained classifier is then encapsulated as a single, powerful LF. Unlike the heuristic-based LFs, which typically vote for one specific intent or abstain, these ML-based LFs can classify a query into \textit{any} of the possible intents, providing broader classification capability.

The combined output of these four strategies is a large set of candidate LFs, as exemplified in Figure~\ref{fig:json_listing_lfs}. This set is then passed to the Pruner for quality control and refinement.

\begin{figure}
\centering
\begin{minipage}{0.91\columnwidth}
\lstdefinestyle{json}{
    basicstyle=\footnotesize\ttfamily,
    % numbers=left,
    % numberstyle=\scriptsize,
    % numbersep=8pt,
    frame=trBL,
    frameround=fttt,
    showstringspaces=false,
    breaklines=true,
    backgroundcolor=\color{white},
    stringstyle=\color{darkblue},
    commentstyle=\color{olive},
    morecomment=[l]{//},
    morestring=[b]",
    morestring=[d]',
    escapeinside={(*@}{@*)},
    tabsize=1
}

\begin{lstlisting}[style=json]
 1  [
 2    {
 3      "(*@\textcolor{purple}{lf\_name}@*)": "LF 1",
 4      "(*@\textcolor{purple}{class\_intent}@*)": "FileCreator",
 5      "(*@\textcolor{purple}{unique\_words}@*)": ["created", "author"],
 6      "(*@\textcolor{purple}{lf\_type}@*)": "ContainsWordLabeller"
 7    },
 8   {
 9      "(*@\textcolor{purple}{lf\_name}@*)": "LF 2",
10      "(*@\textcolor{purple}{class\_intent}@*)": "IssueClosingDate",
11      "(*@\textcolor{purple}{unique\_entities}@*)": ["issue_number"],
12      "(*@\textcolor{purple}{lf\_type}@*)": "EntityLabeller"
13    },
14    {
15      "(*@\textcolor{purple}{lf\_name}@*)": "LF 3",
16      "(*@\textcolor{purple}{class\_intent}@*)": "FileCreator",
17      "(*@\textcolor{purple}{unique\_entities}@*)": ["file_name"],
18      "(*@\textcolor{purple}{unique\_words}@*)": ["author"],
19      "(*@\textcolor{purple}{lf\_type}@*)": "EntityWordLabeller"
20    }
21  ]
\end{lstlisting}
\caption{Example JSON representation of LFs.}
\label{fig:json_listing_lfs}
\end{minipage}
\end{figure}

\subsection{Pruner}
\label{sec:pruner}
The Generator produces a large set of candidate LFs, not all of which are equally effective. Some LFs may have low accuracy, others may have very limited applicability (low coverage), and some may overfit to noisy or overly specific patterns in the training data. The Pruner is the final component of our approach, designed to serve as a quality control filter that evaluates and selects only the most robust and generalizable LFs from this initial candidate pool.

The pruning process begins by holding out a portion of the \textit{expanded dataset} as an evaluation set. Each candidate LF is then applied to this evaluation set to calculate its performance across three key characteristics:
\begin{enumerate}
    \item \textbf{Accuracy:} To measure an LF's labeling accuracy, we use the weighted F1-score \cite{sokolova2009systematic}. This metric is chosen because it provides a balanced measure of precision and recall, which is crucial for multi-class problems.
    \item \textbf{Coverage:} This metric represents the percentage of the evaluation set that an LF labels (i.e., for which it does not abstain). It measures the LF's range of applicability \cite{ratner2020snorkel, ratner2017snorkel}.
    \item \textbf{LF Support:} We introduce this metric to measure the robustness of an LF's underlying pattern. It is defined as the number of queries in the expanded dataset that were used to create the LF. A higher support suggests the LF is based on a more common and reliable signal.
\end{enumerate}

Based on these characteristics, the Pruner executes a multi-step filtering algorithm to construct the final set of LFs. First, to ensure that no intent is ignored, the single best-performing LF (based on F1-score) for each intent class is automatically retained. Second, from the remaining pool of LFs, we apply a core quality filter, discarding any LF that has (a) an accuracy below a predefined threshold, (b) a coverage of zero on the evaluation set, or (c) an LF Support of less than two (i.e., was generated from only a single query). This step removes inaccurate LFs, those with no practical applicability, and those highly likely to have overfitted. 

The final output is a list of high-quality LFs, as shown in Figure~\ref{fig:pruner_output_data}, ready for use in labeling tasks. In our running example, in Figure~\ref{fig:approach}, LF 1 characteristics are higher than the specified threshold for coverage, accuracy, and support. Therefore, it is retained.

\begin{figure}
\centering
\begin{minipage}{0.91\columnwidth}
\lstdefinestyle{json}{
    basicstyle=\footnotesize\ttfamily,
    % numbers=left,
    numberstyle=\scriptsize,
    numbersep=8pt,
    frame=trBL,
    frameround=fttt,
    showstringspaces=false,
    breaklines=true,
    backgroundcolor=\color{white},
    stringstyle=\color{darkblue},
    commentstyle=\color{olive},
    morecomment=[l]{//},
    morestring=[b]",
    morestring=[d]',
    escapeinside={(*@}{@*)},
    tabsize=1
}
\begin{lstlisting}[style=json]
 1  [
 2    {
 3      "(*@\textcolor{purple}{lf\_name}@*)": "LF 1",
 4      "(*@\textcolor{purple}{lf\_type}@*)": "ContainsWordLabeller",
 5      "(*@\textcolor{purple}{coverage}@*)": 0.6,
 6      "(*@\textcolor{purple}{accuracy}@*)": 0.9,
 7      "(*@\textcolor{purple}{lf\_support}@*)": 3
 8    },
 9    {
10      "(*@\textcolor{purple}{lf\_name}@*)": "LF 2",
11      "(*@\textcolor{purple}{lf\_type}@*)": "EntityLabeller",
12      "(*@\textcolor{purple}{coverage}@*)": 0.4,
13      "(*@\textcolor{purple}{accuracy}@*)": 0.8,
14      "(*@\textcolor{purple}{lf\_support}@*)": 2
15    }
16  ]
\end{lstlisting}
\end{minipage}
\caption{Final output of the LFs after the pruner.}
\label{fig:pruner_output_data}
\end{figure}

%% file: Studysetup.tex
The main goal of the proposed approach is to automate the generation of LFs that label users' queries posed to SE chatbots. This section details the SE datasets used to evaluate the effectiveness of the generated LFs. Moreover, it describes the NLU platform used in the evaluation and the configurations employed in our approach for the assessment.

\subsection{Datasets}
\label{sec:datasets}
To evaluate the performance of our approach of generating LFs for SE chatbots, we selected four datasets previously used to develop SE chatbots~\cite{askgitAskGit, lin2020msabot, lei2016semisupervised, abdellatif2021comparison}. Table~\ref{tab:datasets} provides an overview of the number of queries and intents for each of the selected datasets. The selected datasets represent various SE tasks, including seeking information related to software projects and software development tasks. Furthermore, the datasets vary in size, ranging from smaller collections with only a few queries per class to larger sets with hundreds of queries. The details for each dataset with their intents are available in the Appendix. This diversity enables us to assess the effectiveness of our approach across different scales. In particular, we use the following datasets:

\begin{table}
\centering
\caption{Overview of the Selected Datasets Used to Evaluate Our Approach.}
\label{tab:datasets}
\begin{tabular}{@{}l|c|c@{}}
\toprule
\textbf{Dataset} & \textbf{Number of Queries} & \textbf{Number of Intents} \\ 
\midrule
AskGit & 749 & 52 \\ 
MSA & 83 & 8 \\ 
Ask Ubuntu & 50 & 4 \\ 
Stack Overflow & 215 & 5 \\ 
\bottomrule
\end{tabular}
\end{table}

\header{AskGit:}
AskGit~\cite{askgitAskGit} is a chatbot that answers software project-related questions (e.g., ``How many commits happened during March 2021?'') on Slack. It is published on GitHub Marketplace so that practitioners can install it on their software projects. AskGit developers brainstormed to create the initial training set for the intents supported by AskGit. Then, they piloted the chatbot with practitioners to gather additional training queries for each intent, expanding their final dataset. This dataset contains 749 queries grouped into 52 intents.

\header{MSA:}
MSA~\cite{lin2020msabot} is a chatbot that assists practitioners in creating microservice architectures by providing answers to questions about microservice environment settings (e.g., ``Tell me the server's environment setting''). The MSA dataset contains 83 queries across eight distinct intents.

\header{Ask Ubuntu:}
The Ask Ubuntu dataset~\cite{lei2016semisupervised} contains some of the most popular questions from the Ubuntu Q\&A community on Stack Exchange. The intents of the collected questions were annotated through Amazon Mechanical Turk. This dataset includes 50 queries (e.g., ``What screenshot tools are available?'') divided into four intents (e.g., `MakeUpdate').

\header{Stack Overflow:}
Ye et al.~\cite{ye2016software} collected software practitioners' questions posted under the most popular tags on Stack Overflow. Abdellatif et al.~\cite{abdellatif2021comparison} then annotated and categorized these questions (e.g., ``Use of session and cookies what is better?'') into different intents. This dataset contains 215 queries grouped into five intents (e.g., `LookingForBestPractice').

\subsection{NLU Platform}
NLU platforms serve as the backbones for chatbots as they enable chatbots to understand the user's queries~\cite{prajwal2019universal, abdellatif2021comparison, canonico2018comparison}. Typically, chatbot developers resort to off-the-shelf NLUs (e.g., Google Dialogflow) in their chatbot rather than developing an NLU from scratch because it requires both NLP and AI expertise~\cite{abdellatif2021comparison}. Among the variety of NLUs, we select the Rasa NLU platform to evaluate the impact of the generated LFs on the NLU's performance. Our motivation for selecting Rasa is that it is an open-source platform, which makes its internal implementation consistent during our evaluation. Thus, it enables the replicability of our study by other researchers compared to the NLUs on the cloud, whose internal implementations could be changed without any prior notice. Rasa can be installed, configured, and run on local machines, which consumes fewer resources compared to the NLUs that operate on the Cloud. Furthermore, Rasa has been used by prior work to develop SE chatbots~\cite{lin2020msabot, abdellatif2020msrbot, melo2020exploring}.

\subsection{Evaluation Settings}
\label{sec:settings}
Here, we explain the configuration settings used for the evaluation of our approach. For the management and application of LFs, we use \texttt{Snorkel v0.9.8}, which was the latest version at the time the project was initiated. Our approach also involves thresholds for the Generator and Pruner values that influence its performance. 

\begin{itemize}
    \item \textbf{Grouper Threshold:} This threshold determines the minimum semantic similarity score required for an unlabeled query to be added to an existing intent class in the labeled data.
    \item \textbf{Generator Threshold:} This threshold determines the minimum exclusivity score (Equation~\ref{eq:exclusivity}) for a word to be considered exclusive to a particular intent and used in generating an LF.
    \item \textbf{Pruner Threshold:} This threshold determines the minimum accuracy required for an LF to be retained by the Pruner.
\end{itemize}

To determine the optimal values for these thresholds, we conducted a systematic evaluation using the MSRBot dataset~\cite{abdellatif2020msrbot}. We first evaluated the Grouper's threshold, varying from 0.1 to 1.0, while keeping other thresholds constant. We found that a threshold of 0.8 yielded the best overall performance. Next, we evaluated for the Generator's threshold, varying it from 0.1 to 1.0 in increments of 0.1, while keeping the other thresholds constant. We found that a threshold of 0.8 yielded the best overall performance in terms of labeling accuracy. Finally, we evaluated for the Pruner's threshold, also varying each from 0.1 to 1.0 in increments of 0.1, while keeping the other thresholds constant. We found that a threshold of 0.7 yielded the best overall performance in terms of labeling accuracy.

It is important to note that these threshold values, while optimal in the context of our study and the datasets we evaluated, are configurable parameters within our system. Users can adjust these thresholds based on the specific characteristics of their datasets and the desired level of granularity in pattern extraction. This flexibility allows the approach to be tailored to various domains and datasets.

We also set the Pruner's evaluation data size to be 40\% of the expanded labeled data. This value was determined to be optimal through experimentation with different sizes on the MSRBot dataset, with the same increment as the thresholds. Similar to the thresholds, the evaluation data size is a configurable parameter that can be adjusted by the user.

The process of applying labels to data, particularly in situations where different LFs produce conflicting labels, is a critical step in our approach. To systematically manage these conflicts, we adopt Snorkel's Majority Label Voting (MLV) strategy~\cite{ratner2020snorkel}. This choice is informed by manual evaluations and is in alignment with methodologies employed in previous studies~\cite{ratner2017snorkel, lison2020named, lison2021skweak}. The outcome is labeled data ready for NLU training.

\subsection{Performance Evaluation}
To evaluate the performance of the generated LFs and NLU's performance, we also compute the widely used metrics of precision, recall, and F1-score. Precision is the percentage of correctly labeled queries to the total number of labeled queries for that intent (i.e., Precision = $\frac{TP}{TP+FP}$). Recall is the percentage of correctly labeled queries to the total number of queries for that intent in the oracle (i.e., Recall = $\frac{TP}{TP+FN}$). To have an overall performance of the generated LFs, we use the weighted F1-measure that has been used by prior work~\cite{barash2019bridging, ilmania2018aspect}. More specifically, we aggregate the precision and recall using F1-score (i.e., \textit{F1-score} = \(2 \times \frac{Precision \times Recall}{Precision + Recall}\)) for each class and aggregate all classes F1-measure using a weighted average, with the class' support as weights. 

Furthermore, we compute the Area Under the ROC Curve (AUC). AUC assesses the model's ability to distinguish between classes by considering the trade-offs between true and false positive rates~\cite{yang2022learning, bradley1997use}. An AUC value above 0.5 suggests that the model is capable of classifying instances more accurately than random guessing, making it a robust metric for assessing performance in diverse classification scenarios. The significance of AUC is emphasized by its application in prior research, particularly in studies dealing with class imbalance~\cite{bradley1997use, ferri2009experimental, he2009learning, bekkar2013evaluation}, including SE studies that involved analyzing imbalanced datasets~\cite{khatoonabadi2023predicting, khatoonabadi2023wasted, shatnawi2017application}.

Given that our selected datasets involve multiple classes, it is essential to adopt a strategy tailored for multiclass classification. For this purpose, we use the `One vs Rest' (OvR) strategy, which breaks down the multiclass classification problem into individual binary classification tasks, focusing on distinguishing each class against all others. This method allows us to evaluate the model's performance for each class separately and then average these results to obtain an overall performance metric.

%% file: Results.tex
In this section, we present the evaluation results of our proposed approach. For each research question, we provide our motivation, describe the approach to answer the RQ, and discuss the main findings.

\input{RQ1}

\input{RQ2}

%% file: RQ1.tex
\subsection{RQ1: \RQI}
\label{sec:RQ1}
\header{Motivation:}
Prior work shows that manual creation of effective LFs is a tedious, time-consuming task that requires domain expertise~\cite{galhotra2021adaptive}. To alleviate this burden, we propose an automated approach for generating LFs, enabling developers to focus their efforts on the core functionalities of their chatbots rather than on annotating data. Therefore, in this RQ, we evaluate the performance of our generated LFs in labeling users' queries.

\header{Approach:}
To evaluate the performance of our generated LFs, we first randomly split each studied dataset into three distinct sets: labeled, evaluation, and unlabeled data with ratios of 30\%, 20\%, and 50\%. This approach mimics a realistic situation where unlabelled data is more abundant than labeled data, as illustrated by the motivating example in Section~\ref{sec:motivation}. Moreover, this method has been used in similar prior work~\cite{farhour2024weak}. We detail each data split as follows:
\begin{itemize}
    \item \textbf{Labeled Data:} Serves as the basis for creating our LFs. It represents the original data a chatbot would have been trained on. It includes the intent classes with their associated queries (labels).
    \item \textbf{Evaluation Data:} This data is used for evaluating the LFs' labeling performance.
    \item \textbf{Unlabeled Data:} This portion of the data represents the user queries posed to the chatbot, which the developers need to label.
\end{itemize}
We use both the labeled and unlabeled datasets as input to our approach for generating LFs. We reiterate that the Grouper component includes queries from the unlabeled dataset into the labeled set if they are semantically similar, as discussed in Section~\ref{sec:grouper}. In cases where queries have low semantic similarity and are not added to the labeled dataset, we augment them to the evaluation set~\footnote{We refer to the augmented evaluation set as evaluation set.}. We apply the generated LFs to the evaluation set, and a majority-label voting mechanism is used to assign labels to each query. This enables us to evaluate the generated LFs across a more diverse range of queries. We compute the AUC and F1-score by comparing the assigned labels with the ground-truth labels. To minimize bias in our evaluations, we repeated this entire process ten times for each dataset discussed in Section~\ref{sec:datasets}, and we report the average AUC and F1-score across all iterations.

To measure the impact of using the generated LFs on the NLU's performance, we train the NLU on the labeled dataset and augment it with the queries labeled by the generated LFs. We use the evaluation set to assess the NLU's performance. To put our results into perspective, we compare the performance of the NLU trained on data labeled by the generated LFs with two other scenarios: (1) the baseline performance, where the NLU is trained only on the initial labeled data (i.e., labeled data), and (2) a random labeling scenario, where the NLU is trained on the initial labeled data plus additional data with randomly assigned labels. The reason for including the random labeling comparison is to show that the improvement in the chatbot's performance is not merely due to the addition of more training data, but rather the quality of the labels assigned by our LFs.

\header{Results:}
Our approach generates an average number of LFs per run of 288.5 for AskGit, 19.3 for MSA, 15.7 for Ask Ubuntu, and 17.2 for Stack Overflow. The larger AskGit dataset naturally yielded more LFs, while the smaller datasets produced fewer. Table~\ref{tab:auc_score} presents the labeling performance of the generated LFs in terms of AUC and F1-score across the studied datasets. From the table, we observe that the LFs generated by our approach demonstrate promising results in labeling queries in all datasets (except Stack Overflow). Specifically, MSA, AskGit, and Ask Ubuntu exhibit AUC scores above 75\%. AUC scores above 70\% are generally considered good in many machine learning contexts, indicating a strong ability of the LFs to accurately differentiate between classes~\cite{yang2022learning, bradley1997use}. For the Stack Overflow dataset, the generated LFs achieve an AUC score of 51.9\%. To better understand the cause of the low performance on Stack Overflow, we examined the queries with mislabeled intent and found that certain intents in the Stack Overflow dataset have too few queries, making it difficult for our approach to identify distinct patterns. For example, the \textit{FacingError} intent has only 10 queries in the labeled dataset, making it challenging to create an LF that can generalize to other queries for that intent.

\begin{table}
\centering
\caption{Labeling Performance of the Generated LFs.}
\label{tab:auc_score}
\begin{tabular}{@{}l|c|c@{}}
\toprule
\textbf{Dataset} & \textbf{AUC Score (\%)} & \textbf{F1 Score (\%)} \\
\midrule
AskGit & 75.5 & 52.7 \\
MSA & 83.6 & 64.7 \\
Ask Ubuntu & 85.3 & 82.5 \\
Stack Overflow & 51.9 & 44.8 \\
\bottomrule
\end{tabular}
\end{table}

Regarding the impact of using the generated LFs on the NLU's performance, Table~\ref{tab:f1_score} presents the performance comparison between the baseline, our approach, and random labeling for the Rasa NLU in terms of AUC and F1. The difference in percentage points between our approach and the baseline is shown in `Approach Improvement' column, and between our approach and random labeling in `Random Improvement' column. From the table, we observe that training the NLU on data labeled by our approach improves the NLU's performance for all datasets (except Stack Overflow). In the AskGit dataset, our approach resulted in a notable 27.2 percentage point increase in AUC. In the MSA and Ask Ubuntu datasets, the AUC increased by 8.4 and 1.9 percentage points, respectively. On the other hand, the random labeling scenario leads to a decrease in performance across all datasets, with an AUC drop of up to 43.5 percentage points. This underscores the importance of accuracy in data labeling for the NLU's performance. In other words, increasing the training dataset size without ensuring the accuracy of the labels negatively impacts the NLU's performance.

\begin{table*}[t]
\centering
\caption{Comparison of Baseline, Our Approach, and Random Labeling Performance for Rasa NLU.}
\label{tab:f1_score}
\resizebox{\textwidth}{!}{%
\begin{tabular}{@{}l|c c|c c|c c|c c|cc@{}}
\toprule
& \multicolumn{2}{c|}{\textbf{Baseline (\%)}} & \multicolumn{2}{c|}{\textbf{Approach (\%)}} & \multicolumn{2}{c|}{\textbf{Approach Improvement}} & \multicolumn{2}{c|}{\textbf{Random (\%)}} & \multicolumn{2}{c}{\textbf{Random Improvement}}\\ 
\multirow{-2}{*}{\textbf{Dataset}} & \textbf{AUC} & \textbf{F1} & \textbf{AUC} & \textbf{F1} & \textbf{AUC Diff.} & \textbf{F1 Diff.} & \textbf{AUC} & \textbf{F1} & \textbf{AUC Diff.} & \textbf{F1 Diff.}\\ 
\midrule
AskGit & 41.4 & 47.4 & 68.6 & 73.1 & 27.2 & 25.7 & 27.0 & 32.0 & -14.4 & -14.7 \\
MSA & 80.5 & 76.8 & 88.9 & 86.7 & 8.4 & 9.9 & 37.0 & 33.9 & -43.5 & -42.9 \\
Ask Ubuntu & 91.1 & 90.9 & 92.9 & 93.4 & 1.9 & 2.5 & 64.8 & 62.1 & -26.3 & -20.8 \\
Stack Overflow & 41.2 & 60.9 & 35.5 & 55.6 & -5.7 & -5.3 & 34.5 & 49.2 & -6.7 & -11.7 \\
\bottomrule
\end{tabular}%
}
\end{table*}

\begin{tcolorbox}
    \textbf{Summary of RQ1.} Our analysis indicates that LFs generated by our approach generally label data effectively with AUCs of up to 85.3\%. Additionally, when data labeled by our generated LFs are used for training an NLU, they enhance the performance of the chatbot with AUCs of up to 92.9\%.
\end{tcolorbox}

%% file: RQ2.tex
\subsection{RQ2: \RQII}
\header{Motivation:}
The results of RQ1 show that the LFs achieve high performance (AUC $>$ 75\%) in labeling queries across most of the studied datasets. The Pruner component is employed to filter out poorly performing LFs based on specific characteristics discussed in Section~\ref{sec:pruner}. Understanding the impact of these characteristics on LF performance is crucial. This knowledge is essential not only for designing an effective pruning strategy but also for gaining a comprehensive understanding of the LFs' overall impact on labeling performance. In this RQ, we examine the impact of these characteristics (e.g., LF's support) on LF's performance (i.e., AUC).

\header{Approach:}
To investigate the impact of LF characteristics on chatbot performance, we analyze LFs from the perspective of three key characteristics discussed in Section~\ref{sec:pruner}: Coverage, Accuracy, and LF support. For each characteristic, we group LFs based on their performance levels into high, medium, and low-performing categories. Specifically, we rank all generated LFs by each characteristic, selecting the top 20 (high), median 20 (medium), and bottom 20 (low) LFs. Next, we calculate the AUC of each labeled dataset for each characteristic group. We chose groups of 20 LFs to ensure a clear distinction between high, medium, and low-performing groups. Using larger groups diminished the clarity of these distinctions, as the performance characteristics began to overlap between categories. This approach allowed us to maintain a clear separation, ensuring the analysis accurately reflects the distinct impact of LF characteristics on performance without interference from adjacent performance levels.

\begin{table}
    \centering
    \caption{Labeling Performance Across Different LF Characteristics.}
    \label{tab:ablation_study_results}
    \begin{tabular}{@{}c|c|c|c|c@{}}
        \toprule
        \multirow{2}{*}{\textbf{Characteristic}} & \multirow{2}{*}{\textbf{Dataset}} & \multicolumn{3}{c}{\textbf{AUC (\%)}} \\
        & & \textbf{Low} & \textbf{Medium} & \textbf{High} \\
        \midrule
        \multirow{4}{*}{\centering Coverage} & AskGit & 50.5 & 52.3 & 78.2 \\
        & MSA & 60.3 & 78.2 & 86.5 \\
        & Ask Ubuntu & 60.8 & 69.6 & 88.3 \\
        & Stack Overflow & 50.7 & 50.8 & 52.5 \\ 
        \midrule
        \multirow{4}{*}{\centering Accuracy} & AskGit & 50.4 & 51.8 & 52.3 \\
        & MSA & 59.8 & 86.3 & 79.6 \\
        & Ask Ubuntu & 60.7 & 86.3 & 89.3 \\
        & Stack Overflow & 50.6 & 52.5 & 51.6 \\ 
        \midrule
        \multirow{4}{*}{\centering LF Support} & AskGit & 50.4 & 53.5 & 79.4 \\
        & MSA & 60.4 & 86.1 & 86.5 \\
        & Ask Ubuntu & 60.7 & 66.9 & 87.9 \\
        & Stack Overflow & 50.9 & 52.2 & 52.2 \\ 
        \bottomrule
    \end{tabular}
\end{table}
        
\header{Results:}
Table~\ref{tab:ablation_study_results} details the impact of various LF characteristics (i.e., Coverage, Accuracy, and LF Support) on labeling performance, as measured by AUC scores across the studied datasets. Overall, we observe that all characteristics can significantly impact LF performance to varying degrees. In the following, we discuss the impact of each characteristic on the labeling performance.

\header{Coverage:}
From Table~\ref{tab:ablation_study_results}, we observe that LFs with higher coverage consistently achieve better performance compared to those with lower coverage. For example, in the MSA dataset, LFs with high coverage achieve an AUC of 86.5\% compared to 78.2\% for LFs with medium coverage and 60.3\& for LFs with low coverage. This performance improvement is significant for all studied datasets (except Stack Overflow), demonstrating that the proportion of data an LF can label directly impacts its performance.

\header{Accuracy:}
The accuracy has varying effects on labeling performance across different datasets, as shown in Table~\ref{tab:ablation_study_results}. For example, high-accuracy LFs achieve the best performance in the Ask Ubuntu dataset, with an AUC of 89.3\%, while medium-accuracy LFs outperform high-accuracy LFs in the MSA dataset, with an AUC of 86.3\% compared to 79.6\%. Overall, LFs in the high and medium-performing categories outperform those in the low-performing category.

\header{LF Support:}
LFs with more support consistently show better performance, especially in AskGit and Ask Ubuntu, with high AUCs of 79.4\% and 87.9\%, up from 50.4\% and 60.7\%, respectively. Since a higher LF support indicates that the LFs are created with more data points, this result indicates that these LFs are more robust and reliable, leading to improved labeling performance. We also found that larger classes with more queries tend to perform better, as each LF generated for those classes had more data support. For example, in the AskGit dataset, the `number of downloads' class with 32 intents had a median number of intent queries of 4 per LF and a median accuracy of 1. In contrast, the `issue creator' class with 14 intents had a median number of intent queries of 4 per LF and a median accuracy of 0.17.

\smallskip
Our results across the three characteristics reveal two key findings. First, LF characteristics significantly impact performance, with two out of three characteristics (coverage and LF support) showing strong improvements in AUC as their values increase, while accuracy shows varied effects on performance. Second, it is crucial to consider the interplay between LF characteristics when aiming to enhance performance. In other words, focusing on a single characteristic in isolation, such as coverage or accuracy, may lead to some improvements but can also introduce issues. For example, solely prioritizing accuracy may result in LFs with low coverage, as the relationship between coverage and accuracy is not straightforward, as shown in Table~\ref{tab:ablation_study_results}. To measure the correlation between coverage and accuracy, we applied Pearson's correlation test~\cite{cohen2009pearson} on all generated LFs, finding that the correlation varies across datasets, with R values ranging from -0.95 to 0.26, and is significant ($p<.05$) only for AskGit and MSA, where it is negative (-0.35 and -0.95, respectively). To achieve optimal performance across all datasets, it is essential for the practitioners to adopt a balanced approach that takes into account all LF characteristics collectively.

\begin{tcolorbox}
    \textbf{Summary of RQ2.} Our findings indicate that higher values in LF characteristics generally correlate with improved labeling performance. While the influence of characteristics on performance varies, focusing solely on one characteristic may negatively impact others, suggesting a balanced approach that considers all characteristics is essential for optimal performance.
\end{tcolorbox}

%% file: Discussion.tex
Our evaluation shows that the proposed approach effectively generates LFs for labeling SE data. In this section, we discuss two important considerations for practitioners. First, we examine the relevance of NLU-based chatbots in the era of LLMs, as this context is crucial for understanding where our approach fits in the current landscape of conversational AI. Second, we analyze how the number of generated LFs affects labeling performance, since our approach can produce hundreds of LFs (e.g., 288.5 per run for AskGit) and applying all of them has computational costs. These discussions provide practical insights for deploying our approach in real-world SE chatbot development.

\subsection{NLU-based Chatbots in the Era of LLMs}
\label{sec:D1}
NLU platforms (e.g., Rasa and Dialogflow) have been the backbone of task-oriented chatbots, especially in enterprise settings. These chatbots rely on intent classifiers and entity extractors to understand user input and respond appropriately \cite{abdellatif2020msrbot}. However, the emergence of LLMs has transformed various domains such as SE and customer services. LLMs can engage in open-ended dialogue by leveraging massive pretrained knowledge. This raises a key question: ``Are traditional NLU-based chatbots still relevant for building chatbots (particularly in SE domains), or are LLMs poised to replace them?''. Recent research suggests that while LLMs dramatically expand what chatbots can do, NLU frameworks remain valuable due to their reliability, control, and efficiency \cite{mctear2023comparative, foosherian2023enhancing}. For example, McTear et al. \cite{mctear2023comparative} show that NLU pipelines still outperform LLMs on safety-critical, database-lookup tasks, while Foosherian et al. \cite{foosherian2023enhancing} demonstrate that a pipeline provides guard-rails that keep GPT-4’s powerful but non-deterministic outputs in check. Together, these studies underline that an NLU platform offers deterministic execution, transparent debugging, and low-latency inference, attributes that remain indispensable even as LLMs are layered on top. Below, we begin by revisiting our focus on data labeling for SE chatbots to clarify the relationship between our approach and the evolving landscape of AI models for conversational systems. We then review the technical capabilities of each approach, industry adoption trends, cost-effectiveness considerations, and other factors (like resource constraints, context handling, and efficiency) to better understand their roles in modern SE chatbots.

\subsubsection{Focus on Data Labeling}\hfill

Our primary contribution is an approach for automatically generating LFs that can efficiently label training data for SE chatbots. This labeled data could be used for various applications, including both training specialized NLU platforms and fine-tuning LLMs \cite{abdellatif2021comparison, liu2024understanding}. In our evaluation, we demonstrate the effectiveness of our LF-labeled data by training an NLU-based chatbot, as this provides a clearer demonstration of the impact of our labeled data compared to LLMs. This is because, unlike LLMs, which are pre-trained on vast corpora that include SE-related content \cite{brown2020language, liu2024understanding}, NLUs start with minimal domain-specific knowledge \cite{abdellatif2021comparison}. This creates a controlled environment where performance improvements can be directly attributed to our LF-labeled data. With LLMs, any performance gains might be obscured by the model's pre-existing knowledge, making it difficult to isolate the contribution of our labeled data.

\subsubsection{Why are NLU-based Chatbots still needed}\hfill

\noindent \textbf{Architectural Differences.} Traditional NLU-based conversational chatbots use a pipeline architecture with explicit modules for NLU (intent classification and entity recognition), dialogue management (state tracking and policy), and NLG (response templates or generation) \cite{foosherian2023enhancing}. For example, a Rasa or Dialogflow assistant will parse an input (“I lost my password”) into a structured intent (e.g., report\_issue) with entities (e.g., issue\_type=password), then decide the next action via defined rules or a learned policy, and finally produce a response (often from a template or small model). These NLU-based systems require developers to predefine intents, entities, and sometimes dialogue logic for the task domain \cite{mctear2023comparative}. 

\noindent \textbf{Structured Control.} NLU-based chatbots operate within a constrained, known action space. The dialogue manager will only execute explicitly defined actions or give responses written by chatbot developers. This makes them more predictable and easier to control \cite{foosherian2023enhancing}. Errors are often systematic (e.g., misclassification of an intent) and can be debugged by adjusting training data or rules. As one study notes, the pipeline approach grants designers transparency and control over the agent’s behavior, helping counteract issues like hallucinations and enabling explainability for decisions \cite{foosherian2023enhancing}. For example, a banking chatbot built with an NLU pipeline will never output an off-topic or fabricated answer---if it does not understand, it might default to a fallback message. This reliability is crucial in many SE chatbot scenarios (e.g., a DevOps assistant executing deploy commands should only do what it is programmed to do).

\noindent \textbf{Reliability and Truthfulness.} LLMs, while powerful, may hallucinate, generating responses that seem plausible but are actually incorrect \cite{foosherian2023enhancing}. 
For example, an LLM-based chatbot might fabricate an API name or misstate a library’s capabilities when assisting a programmer, whereas an NLU-based chatbot would simply say it cannot help if it does not have that knowledge or has low confidence in its response \cite{abdellatif2021comparison}. It has been shown that LLM responses are not always reliable and can be factual but incorrect in subtle ways \cite{foosherian2023enhancing}. In an SE setting, this unreliability could mean suggesting insecure code or wrong technical advice – indeed, one study noted LLMs can output ``harmful advice, buggy code, or inaccurate information'' without realizing it \cite{bocklisch2024task}. This is a serious concern for professional use. Additionally, LLMs lack guaranteed consistency: the same question phrased differently might yield slightly different answers \cite{wang2024prompt, zhuo2024prosa}, and they do not inherently follow a predefined style or persona (unless controlled via prompting) \cite{frisch2024llm, choi2025examiningidentitydriftconversations}. Maintaining a consistent tone or enforcing business-specific messaging is more straightforward with template-based NLG in an NLU system, whereas with LLMs it requires careful prompt engineering or fine-tuning \cite{asai2020logic, karanikolas2023large, mctear2023comparative}.

\noindent \textbf{Data and Learning Dependencies.} While LLMs can adapt to tasks through in-context learning and fine-tuning, both methods present challenges for domain-specific applications \cite{dong2022survey}.
Furthermore, improving domain-specific accuracy through fine-tuning is particularly challenging due to the lack of SE benchmarks and substantial annotated datasets \cite{shin2023prompt}, which reinforces the need for efficient data-labeling methods like ours.

\noindent \textbf{Security and Privacy.} A significant consideration, especially for SE chatbots handling proprietary code or internal data, is security and privacy. Using cloud-hosted LLM APIs (like OpenAI's) can pose risks if sensitive information is inadvertently sent to third-party servers \cite{Ray_2024}. While enterprise tiers offer better privacy guarantees (e.g., no data retention for training) \cite{OpenAI_2023}, many organizations, particularly in regulated sectors like finance or healthcare, prefer or require solutions that keep data entirely on-premises. Traditional NLU frameworks like Rasa can be easily deployed on-premises, offering full data control, unlike larger LLM models, which are computationally intensive to host and scale locally.

\subsection{Impact of the Number of LFs on Labeling Performance}
\label{sec:D2}
We found that the LFs generated by our approach effectively label data, as discussed in RQ1. However, there is no such thing as a free lunch. Using our approach requires applying each generated LF to all user queries, which can be time and energy consuming. Therefore, in this section, we investigate the impact of the number of LFs on labeling performance. To accomplish this, we progressively assess the impact of adding LFs on labeling performance. Specifically, we construct a set of LFs by randomly selecting one LF from the LFs generated by our approach in RQ1 and adding it to the set. The set contains all generated LFs, averaging 288.5 per run for AskGit, 19.3 for MSA, 15.7 for Ask Ubuntu, and 17.2 for Stack Overflow. These LFs are randomly shuffled. Next, we sequentially select the next LF from the shuffled list, apply it to the queries in the evaluation set, and repeat the evaluation process to measure the impact of adding more LFs on labeling performance. We continue this process until all LFs in the list have been applied.

\begin{figure}
	\centering
	% \hspace{-0.1\linewidth} % Shifts the figure to the left by 10% of the line width
	\includegraphics[width=.8\linewidth]{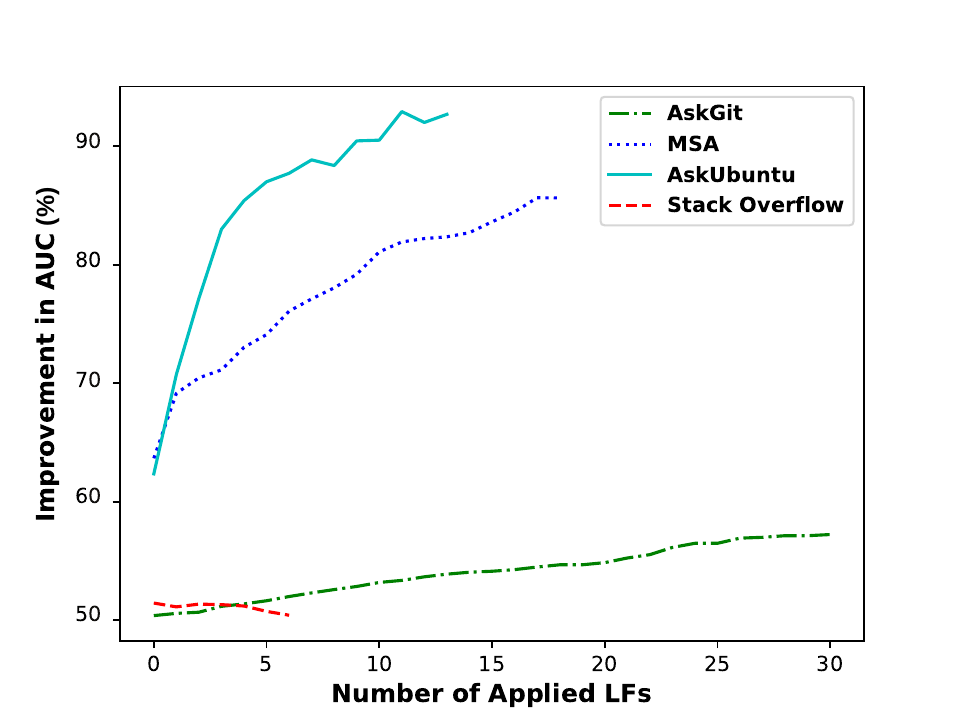}	
	\caption{Impact of the number of pruned LFs on labeling performance.}
	\label{fig:number_of_lfs}	
\end{figure}

Figure~\ref{fig:number_of_lfs} shows the performance in terms of AUC when applying LFs additively. From the figure, we observe that performance increases as more LFs are applied across all datasets (except Stack Overflow), though with varying magnitudes. For example, in the Ask Ubuntu dataset, we observe a gradual increase in labeling performance as more LFs are added, reaching a peak AUC of 92\%. Similarly, in the MSA and AskGit datasets, labeling performance increases as more LFs are applied, reaching up to 85\% and 75\%, respectively. We truncated the AskGit dataset at 30 LFs in Figure~\ref{fig:number_of_lfs} for visual clarity. Nevertheless, the results from the AskGit dataset show that labeling performance continues to improve as more LFs are applied (beyond the 30 LFs), reaching an AUC of 75\% when all LFs in the shuffled list are applied. The Stack Overflow dataset deviates from the general trend, with its performance remaining relatively constant around 50\% AUC, regardless of the number of LFs applied. This deviation can be attributed to the specific challenges outlined in RQ1, such as the lack of LFs for certain intent classes.

Another interesting observation is that the performance saturated when a high number of LFs were applied. Upon closer examination of the results, we find that performance improves incrementally as more LFs are added, continuing until all intents in the dataset are adequately covered by the applied LFs, after which the performance gain plateaus. Our findings underscore the importance of applying more LFs that cover various intents, rather than focusing on a single intent. That said, users of our approach should consider the characteristics of the chatbot. For example, in a chatbot that refactors code, it is expected that the majority of user queries will focus on refactoring rather than unrelated topics like the weather. In this case, more LFs targeting refactoring questions should be applied.

%% file: Threats.tex
In this section, we discuss the threats to the internal and external validity of our study.

\header{Threats to Internal Validity.}
Internal validity concerns factors that could have influenced our results. The choice of the threshold in our evaluation might influence the results. Using a different threshold may yield different outcomes. To mitigate this threat, we experimented with various threshold values using the MSRBot dataset~\cite{abdellatif2020msrbot}, which is not included in our evaluation, as discussed in Section~\ref{sec:settings}. MSRBot has a sufficient number of intents and queries and has been used in prior SE work~\cite{abdellatif2021comparison, abdellatif2024transformer}. We selected the best-performing thresholds based on a manual examination of the results. Another threat is the choice of tool used for detecting entity types in the LF Generator component. To mitigate this, we relied on Rasa which has previously been shown to achieve good performance in extracting SE entities~\cite{abdellatif2021comparison}.

\header{Threats to External Validity.}
External validity concerns the generalizability of our findings. We evaluated the LFs generated by our approach using the AskGit, MSA, Ask Ubuntu, and Stack Overflow tasks. Hence, our findings may not generalize to other tasks within the SE domain. However, we believe that these datasets cover very common SE tasks such as gathering information related to software projects (e.g., AskGit) and seeking information about operating systems (e.g., Ask Ubuntu), which could be improved with chatbots. That said, we encourage other researchers to replicate our study by considering additional SE datasets.

To evaluate the impact of training the NLU using labeled queries on labeling performance, we conducted a case study using Rasa, as discussed in RQ1. This may affect the generalizability of our results. However, Rasa is widely used by chatbot developers and researchers to develop SE chatbots. Additionally, Rasa is an open-source NLU, meaning its internal implementation remains fixed, unlike closed-source NLUs where the internal implementation might change without prior notice to users. This stability enables other practitioners to replicate our study. Finally, the results show that our approach effectively labels user queries correctly. Therefore, training the NLU on high-quality labeled data leads to improved performance.

%% file: Relatedwork.tex
This paper introduces an approach for automating the generation of LFs for SE chatbots. Accordingly, we will explore two relevant areas in the related works: firstly, the existing literature on SE chatbots, and secondly, the studies in data labeling.

\subsection{SE Chatbots} 
Chatbots have been developed and extensively studied across various domains, including education~\cite{wollny2021there}, healthcare~\cite{parviainen2022chatbot}, and customer service~\cite{misischia2022chatbots}. In the field of SE, they have been employed for a range of purposes where they have had a significant impact, such as assisting developers in task automation, providing guidance to newcomers, and facilitating information retrieval from software repositories~\cite{romero2020experiences, dominic2020conversational, abdellatif2020msrbot, lin2020msabot, zhang2020chatbot4qr, bradley2018context, serban2021saw}.

Dominic et al.~\cite{dominic2020conversational} developed a chatbot using Rasa to assist newcomers in the onboarding process to open source projects, offering guidance, resources, and mentor recommendations. The developed chatbot helps integrate newcomers into the community more effectively. Abdellatif et al.~\cite{abdellatif2020msrbot} introduced MSRBot, a chatbot that answers questions extracted from software repositories, such as identifying commits that fix specific bugs. This chatbot significantly enhances the accessibility of information in software repositories. Lin et al.~\cite{lin2020msabot} leveraged Rasa to create MSABot, a chatbot framework aimed at supporting the development and operation of microservice-based systems, addressing challenges like modularization and scalability. Finally, Bradley et al.~\cite{bradley2018context} proposed Devy, a context-aware conversational assistant that streamlines the development process by reducing manual low-level command execution, allowing developers to focus on high-level tasks.

The increased attention to SE chatbots and the challenges associated with collecting data to train them~\cite{santhanam2022bots, dominic2020conversational, abdellatif2020msrbot} serve as the motivation for our study. Our aim is to assist practitioners in improving chatbot performance in intent classification while reducing resource costs by automating the annotation of user input. Our study differs in that we focus on supporting chatbot practitioners rather than developing chatbots ourselves.

\subsection{Data Labeling}
There has been a substantial quantity of work on data labeling in recent years~\cite{varma2018snuba,li2021weakly,zhao2021glara,hancock2018training}. For example, Li et al.~\cite{li2021weakly} developed a method that uses a few initial rules as a starting point to identify and classify text segments, combining these rules with machine learning models. Zhao et al. introduced GLARA~\cite{zhao2021glara}, which utilizes graph-based techniques to expand and refine rules for naming and categorizing entities in texts. Hancock et al. ~\cite{hancock2018training} use natural language explanations to automatically generate LFs. This method allows users to explain their reasoning in simple language, which the system then translates into rules for data classification. Finally, Boecking et al. developed a framework~\cite{boecking2020interactive} that enhances label generation by actively learning from user feedback. This interactive approach allows the system to continuously improve its labeling accuracy through user-guided heuristics. The work closest to ours is the work by Varma et al.~\cite{varma2018snuba}, which proposed an approach that automates the creation of ML-based heuristics. Varma et al. evaluated their approach on medical, hardware, and text classification datasets including tasks such as image classification (bone tumor, mammogram) and sentiment analysis (Twitter sentiments).

To the best of our knowledge, there is no existing work that proposes an approach specifically tailored to improving NLU performance for SE chatbots. Our work differs from and complements prior efforts in three ways. First, our approach is a fully automated end-to-end process for generating LFs for SE chatbots. Second, our approach utilizes various characteristics of user queries—such as unique words, entity types, and distinct combinations---going beyond solely ML-based LFs. Third, we examine the specific characteristics of generated LFs that contribute most to their effectiveness, providing insights to improve LF quality and overall labeling performance. We believe our work complements prior work in generating LFs for the SE domain. Furthermore, we analyze the impact of the number of LFs on labeling performance, providing practical guidance for optimizing LF usage.

%% file: Conclusion.tex
In this paper, we addressed the challenge of efficiently labeling data for SE chatbots by proposing an automated approach for generating LFs. Our approach extracts patterns from a set of labeled data and uses those patterns to generate LFs capable of labeling a larger, unlabeled dataset. We evaluated the effectiveness of the generated LFs on four diverse SE datasets and found that they performed well in labeling tasks, demonstrating their ability to capture domain-specific knowledge. Furthermore, we trained an NLU chatbot using the auto-labeled data and observed performance improvements in most cases, confirming the high quality of the generated LFs. We also investigated the characteristics of LFs that influence their performance and discovered that increasing values of coverage, accuracy, and LF support generally lead to better labeling performance, albeit to varying degrees across datasets. Our approach contributes to the field of SE chatbots by automating the data labeling process, allowing chatbot practitioners to focus on core functionalities and accelerate the development process.

%% file: Appendix.tex
\section{Supplementary Data}
\label{appendix:supplementary}

% Table for MSA dataset
\begin{table}[ht]
\caption{Intents, Query Distribution and Definitions in the MSA Dataset}
\label{tab:msa_intents}
\centering
\resizebox{.65\textwidth}{!}{%
\begin{tabular}{@{}l|c|p{6cm}@{}}
\toprule
\textbf{Intent} & \textbf{\# Queries} & \textbf{Definition} \\
\midrule
service\_using\_info & 14 & Provides usage information or amount overview for a specific service. \\
service\_info & 12 & Provides information or details about a specific service. \\
service\_only & 12 & Handle queries that mention only the service name, possibly providing general information. \\
service\_api\_list & 11 & Provides a list of APIs for a specific service. \\
service\_health & 10 & Provides health data or status for a service or server. \\
service\_dependency\_graph & 9 & Provides the dependency graph or dependency information between services. \\
service\_env & 8 & Provides environment settings or information about the server environment. \\
last\_build\_fail & 6 & Provides information or reasons about the last build failure for a service. \\
\bottomrule
\end{tabular}
}
\end{table}

% Table for Ask Ubuntu dataset
\begin{table}[ht]
\caption{Intents, Query Distribution and Definitions in the Ask Ubuntu Dataset}
\label{tab:askubuntu_intents}
\centering
\resizebox{.65\textwidth}{!}{%
\begin{tabular}{@{}l|c|p{6cm}@{}}
\toprule
\textbf{Intent} & \textbf{\# Queries} & \textbf{Definition} \\
\midrule
Software Recommendation & 17 & Provides recommendations for software based on user needs. \\
Shutdown Computer & 13 & Provides guidance or commands on how to shut down the computer. \\
Make Update & 10 & Provides instructions or information on how to update or upgrade Ubuntu versions. \\
Setup Printer & 10 & Provides instructions on how to set up printers on Ubuntu. \\
\bottomrule
\end{tabular}
}
\end{table}

% Table for Stack Overflow dataset
\begin{table}[ht]
\caption{Intents, Query Distribution and Definitions in the Stack Overflow Dataset}
\label{tab:stackoverflow_intents}
\centering
\resizebox{.65\textwidth}{!}{%
\begin{tabular}{@{}l|c|p{6cm}@{}}
\toprule
\textbf{Intent} & \textbf{\# Queries} & \textbf{Definition} \\
\midrule
LookingForCodeSample & 132 & Requests code samples or examples to solve a programming problem. \\
UsingMethodImproperly & 51 & Requests help for incorrect usage of a method or function. \\
LookingForBestPractice & 12 & Seeks best practices or coding recommendations. \\
PassingData & 10 & Inquires about passing data between components or functions. \\
FacingError & 9 & Requests help to resolve an error or exception encountered. \\
\bottomrule
\end{tabular}
}
\end{table}

% Table for AskGit dataset
\begin{table*}[ht]
\centering
\caption{Intents, Query Distribution and Definitions in the AskGit Dataset}
\label{tab:askgit_intents}
\resizebox{\textwidth}{!}{%
\begin{tabular}{@{}l|c|p{10cm}@{}}
\toprule
\textbf{Intent} & \textbf{\# Queries} & \textbf{Definition} \\
\midrule
number\_of\_downloads & 32 & Retrieves the total number of downloads for the repository. \\
list\_collaborators & 30 & Provides a list of collaborators or contributors to the repository. \\
number\_of\_subscribers & 29 & Retrieves the total number of subscribers to the repository. \\
number\_of\_forks & 29 & Retrieves the total number of forks of the repository. \\
number\_of\_stars & 29 & Retrieves the total number of stars of the repository. \\
number\_of\_commits & 29 & Retrieves the total number of commits in the repository. \\
number\_of\_branches & 28 & Retrieves the total number of branches in the repository. \\
issue\_related\_commits & 22 & Retrieves the commits that are related or linked to a specific issue. \\
list\_branches & 22 & Provides a list of all branches in the repository. \\
file\_creator & 21 & Identifies the creator or initial developer of a specific file. \\
repository\_creation\_date & 19 & Provides the creation date of the repository. \\
repository\_topics & 18 & Provides the topics associated with the repository. \\
issue\_contributors & 17 & Lists the developers who contributed to a specific issue. \\
repository\_license & 16 & Provides the license information of the repository. \\
issue\_closer & 15 & Identifies the developer who closed or resolved a specific issue. \\
number\_of\_watchers & 15 & Retrieves the total number of watchers of the repository. \\
default\_branch & 15 & Provides the name of the default branch of the repository. \\
main\_programming\_language & 15 & Provides the main programming language used in the repository. \\
repository\_owner & 15 & Identifies the owner of the repository. \\
issue\_creator & 14 & Identifies the developer who created or opened a specific issue. \\
issue\_closing\_date & 14 & Provides the closing date of a specific issue. \\
pr\_closer & 14 & Identifies the developer who closed or merged a specific pull request. \\
pr\_creator & 14 & Identifies the developer who created or opened a specific pull request. \\
pr\_contributors & 14 & Lists the developers who contributed to a specific pull request. \\
most\_recent\_issues & 13 & Provides a list of the most recent issues in the repository. \\
top\_contributors & 13 & Lists the top contributors to the repository. \\
pr\_closing\_date & 13 & Provides the closing date of a specific pull request. \\
most\_recent\_prs & 12 & Provides a list of the most recent pull requests in the repository. \\
files\_changed\_by\_pr & 12 & Lists the files that were changed by a specific pull request. \\
number\_of\_issues & 12 & Retrieves the total number of issues in the repository. \\
pr\_creation\_date & 11 & Provides the creation date of a specific pull request. \\
initial\_commit & 11 & Provides information about the initial or first commit in the repository. \\
issue\_creation\_date & 11 & Provides the creation date of a specific issue. \\
number\_of\_prs & 11 & Retrieves the total number of pull requests in the repository. \\
activity\_report & 10 & Provides a report of recent repository activity. \\
longest\_open\_issue & 10 & Identifies the issue that has been open for the longest time. \\
longest\_open\_pr & 9 & Identifies the pull request that has been open for the longest time. \\
latest\_commit & 9 & Provides information about the latest or most recent commit in the repository. \\
latest\_commit\_in\_branch & 9 & Provides information about the latest commit in a specific branch. \\
latest\_release & 9 & Provides information about the latest release in the repository. \\
largest\_files & 9 & Identifies the largest files in the repository, by size or lines of code. \\
commits\_in\_pr & 8 & Lists the commits included in a specific pull request. \\
list\_languages & 8 & Lists the programming languages used in the repository. \\
number\_of\_collaborators & 8 & Retrieves the total number of collaborators or contributors to the repository. \\
contributions\_by\_developer & 8 & Provides the number of contributions made by a specific developer. \\
initial\_commit\_in\_branch & 8 & Provides information about the initial or first commit in a specific branch. \\
pr\_assignees & 7 & Identifies the assignees of a specific pull request. \\
number\_of\_commits\_in\_branch & 7 & Retrieves the number of commits in a specific branch. \\
last\_developer\_to\_touch\_a\_file & 7 & Identifies the last developer who modified a specific file. \\
issue\_assignees & 6 & Identifies the assignees of a specific issue. \\
developers\_with\_most\_open\_issues & 6 & Lists developers who have the most open issues assigned to them. \\
list\_releases & 6 & Provides a list of releases in the repository. \\
\bottomrule
\end{tabular}
}
\end{table*}